\documentstyle[12pt]{article}
\normalbaselines

\begin{document}

\baselineskip=24pt

\bibliographystyle{unsrt}
\vbox {\vspace{6mm}}
\bigskip
\bigskip
\bigskip
\bigskip
\begin{center} {\bf CLASSICAL-LIKE DESCRIPTION OF QUANTUM DYNAMICS\\
BY MEANS OF SYMPLECTIC TOMOGRAPHY}
\end{center}

\begin{center} 
{\it Stefano Mancini, Vladimir I. Man'ko\footnote{On leave from Lebedev 
Physical Institute, Moscow}
 and Paolo Tombesi }
\end{center}

\begin{center}
Dipartimento di Matematica e Fisica, Universit\`a di Camerino, I-62032 
Camerino
\end{center}

\bigskip
\bigskip
\bigskip

\begin{abstract}
The dynamical equation of quantum mechanics are rewritten in form of 
dynamical equations for the
measurable, positive marginal distribution of the shifted, rotated and 
squeezed quadrature
introduced in the so called "symplectic tomography". Then the possibility 
of a purely classical
description of a quantum system as well as a reinterpretation of the 
quantum measurement theory is
discussed and a comparision with the well known quasi-probabilities 
approach is given.
Furthermore, an analysis of the properties of this marginal distribution, 
which contains all the
quantum information, is performed in the framework of classical probability 
theory. Finally
examples of harmonic oscillator's states dynamics are treated.  

\end{abstract}

PACS Number(s): 03.65.Bz, 03.65.Ca, 42.50.Dv

\bigskip
\bigskip
\bigskip
\bigskip

\section{Introduction}\label{s1}

".....Schr\"odinger made no secret of his intention to substitute 
simple classical pictures for the
strange conceptions of quantum mechanics, for whose abstract 
character he expressed deep aversion".
It is clear from this commentary of Rosenfeld \cite{Ros} that 
from the early days of quantum theory
there has been a permanent wish to understand 
quantum mechanics in terms of classical probabilities. However,
due to the Heisemberg \cite{Heisemberg} and Schr\"odinger-Robertson 
\cite{Schrodinger},
\cite{Robertson} uncertainty relation for the position and momentum 
in quantum systems, does not
exist joint distribution function in the phase space. 
This leads to the introduction of the so called
quasi-probability distributions, such as Wigner function \cite{Wigner}, 
Husimi Q-function
\cite{Husimi} and Glauber-Sudarshan P-function \cite{Glauber}, 
\cite{Sudarshan}; later on 
unified into one-parametric family \cite{CahillGlauber}. 
Furthermore, in order to get a
bridge between quantum and classical physics, Madelung
\cite{Madelung} already observed that the modulus and the phase 
of the wave function obey the
hydrodinamical classical equations, and along this line the stochastic 
quantization scheme has been
suggested by Nelson
\cite{Nelson} to link the classical stochastic mechanics formalism with 
the quantum mechanical
basic entities, such as wave function and propagator.   In
some sense, also the hidden variables \cite{Belinfante} was proposed 
to relate the quantum
processes to the classical ones. Nevertheless, up to date does not exsist 
a formalism which
consistently connects the "two worlds". 

The discussed quasi-probabilities illuminated the similarities and
the differences between classical and quantum considerations, and they 
are widely used as
instruments for calculations in quantum theory \cite{Scully}, 
\cite{Tatarsky}. However, they cannot
play the role of classical distributions since, for example, 
the Wigner function and the
P-function may have negative values. Althought the Q-function 
is always positive and
normalized, it does not describe measurable distributions of 
concrete physical variables. 

Recentely, after J. Bertrand and P. Bertrand \cite{Bert} made 
the first attempt to apply the
tomographic principle to phase space distributions, 
Vogel and Risken \cite{VogelRisken},
using the formalism of Ref.
\cite{CahillGlauber}, established an integral relation between the
Wigner function and the marginal distribution for the measurable 
homodyne output variable which
represents a rotated quadrature of the electromagnetic field. 
This result gives the possibility of
"measuring" the quantum state, and it is referred as optical 
homodyne tomography \cite{Raymer}. 

In Ref. \cite{Man1} a symplectic tomography procedure was 
suggested to obtain the Wigner
function by measuring the marginal distribution for a shifted, 
rotated and squeezed quadrature, 
which
depends on extra parameters. In Ref. \cite{DarPaul} the 
formalism of Ref. \cite{VogelRisken}
was formulated in invariant form, relating the homodyne 
output distribution directly to the
density operator. Then, in Ref. \cite{Man2} the symplectic 
tomography formalism was also formulated
in this invariant form and it was extended to the multimode 
case. Thus, due to the introduction
of quantum tomography procedure the real positive marginal 
distribution for measurable
observables, such as rotated shifted and squeezed quadratures, 
turned out to determine
completely the quantum states.

The aim of the present work is to formulate the standard quantum 
dynamics in terms of the classical
marginal distribution of the measurable shifted, rotated and 
squeezed quadrature components, used in
the symplectic tomography scheme. Thus we obtain an alternative 
formulation of the quantum system
evolution in terms of evolution of a real and positive  distribution 
function for measurable
physical observables. We will show the connection of such a "classical" 
probability evolution
with the evolution of the above discussed quasi-probability distributions.
Preliminarly, the approach was shortly presented in Ref. \cite{PLA}.

Examples relative to states of harmonic oscillator and free motion 
will be considered in
the frame of the given formulation of quantum mechanics as well as 
oscillator with friction and
driven terms included.

\section{Density operator and distribution for shifted rotated and 
squeezed quadrature}

In Ref. \cite{Man1} it was introduced an operator $\hat X$ as the
generic linear combination of the position $\hat q$ and momentum 
$\hat p$ ($\hbar=1$)
\begin{equation}\label{X}
\hat X=\mu \hat q+\nu\hat p+\delta,
\end{equation}
which depends upon three real parameters $\mu$, $\nu$, $\delta$ and, 
due to its hermiticity, is a
measurable observable. Thus, the marginal distribution, defined as 
the Fourier transform of the
characteristic function 
\begin{equation}
w(X,\mu,\nu,\delta)=\int\; dk\,e^{-ikX}\langle e^{ik{\hat X}}\rangle\,,
\end{equation}
depends itself upon the parameters
$\mu,\nu,\delta$, and it is 
normalized with respect to
the $X$ variable
\begin{equation}\label{norm}
\int \;dX\,w(X,\mu,\nu,\delta)=1\,.
\end{equation}
Furthermore, it was shown \cite{Man1} that this marginal distribution
is related to the state of the quantum system, expressed in terms of 
its Wigner
function $W(q,p)$, as follows
\begin{equation}\label{wX}
w(X,\mu,\nu,\delta)=\int e^{-ik(X-\mu q-\nu p-\delta)}W(q,p)
\frac{dkdqdp}{(2\pi)^2}.
\end{equation}
Eq. (\ref{wX}) shows that $w$ is a function of the difference 
$X-\delta=x$, so that it can be
rewritten as 
\begin{equation}\label{w}
w(x,\mu,\nu)=\int e^{-ik(x-\mu q-\nu p)}W(q,p)\frac{dkdqdp}{(2\pi)^2}.
\end{equation}

This formula can be inverted and the Wigner function of the state 
can be expressed in
terms of the marginal distribution \cite{Man1}
\begin{equation}\label{W}
W(q,p)=(2\pi)^2z^2w_F(z,-zq,-zp),
\end{equation}
where $w_F(z,a,b)$ is the Fourier component of the marginal 
distribution (\ref{w}) taken with
respect to the variables $x,\mu,\nu$, i.e.
\begin{equation}\label{wF}
w_F(z,a,b)=\frac{1}{(2\pi)^3}\int w(x,\mu,\nu)e^{-i(xz+\mu a+\nu b)}
dxd\mu d\nu.
\end{equation}
Hence, it was shown that the quantum state could be described by 
the positive classical marginal
distribution for the squeezed, rotated and shifted quadrature which 
could be considered as a
classical probability associated to a stochastic variable $x$ and 
depending also on parameters.

In the case of only rotated
quadrature, $\mu=\cos\phi$ and $\nu=\sin\phi$, the usual optical 
tomography formula of
Ref.
\cite{VogelRisken}, gives the same possibility  through the Radon 
transform instead of the Fourier
transform. This is, in fact, a partial case of the symplectic 
transformation of quadrature since
the rotation group is a subgroup of the symplectic group
$ISp(2,R)$ whose parameters are used to describe the transformation 
(\ref{X}).

In Ref. \cite{Man2} an invariant form connecting directly the 
marginal distribution
$w(x,\mu,\nu)$ and the density operator was found
\begin{equation}\label{rho}
\hat\rho=\int dxd\mu d\nu\;w(x,\mu,\nu) \hat K_{\mu,\nu},
\end{equation}
where the kernel operator has the form
\begin{equation}\label{K}
\hat K_{\mu,\nu}=\frac{1}{2\pi}z^2e^{-izx}
e^{-iz^2\mu\nu/2}
e^{iz\nu \hat p}
e^{iz\mu \hat q}.
\end{equation}
The formulae (\ref{W}) and
(\ref{rho}) of symplectic tomography show that there exist an 
invertible map between the quantum
states described by the set of nonnegative and normalized 
hermitian density operators $\hat\rho$
and the set of positive, normalized marginal distributions 
("classical" ones) for the measurable
shifted, rotated and squeezed quadratures. So, the information 
contained in the marginal
distribution is the same which is contained in the density operator; 
and due to this, one could
represent the quantum dynamics in terms of evolution of the marginal 
probability.
Really, the fact that $\hat K_{\mu,\nu}$ depends also on the $z$ 
variable (i.e. each Fourier
component gives a selfconsistent kernel) shows the overcompleteness 
of information achievable by
measuring the observable of Eq. (\ref{X}).

The definition of the marginal distribution function $w(x,\mu,\nu)$ 
might be alternatively
given in terms of the eigenstates of the operator $\hat x=\hat X-\delta$
\begin{equation}\label{Xeigen}
\hat x|x\rangle=x|x\rangle
\end{equation}
which can be obtained from the position eigenstates
\begin{equation}\label{qeigen}
\hat q|q\rangle=q|q\rangle
\end{equation}
by the action of the unitary operator $\hat S$
\begin{equation}\label{S}
|x\rangle=\hat S|q\rangle
\end{equation}  
which represents the composition of simple operations such as rotation 
and squeezing, i.e. it 
satisfies the requierement
\begin{equation}\label{Saction}
{\hat S}^{\dag}\hat q\hat S=\mu\hat q+\nu \hat p.
\end{equation}
It is worth to remark, about this transformation, that there exist 
a costraint \cite{Yuen}
due to the commutation relation between the observable (\ref{X}) 
and its canonical
conjugate, i.e. if one introduce the observable 
\begin{equation}
P=\mu'\hat q+\nu'\hat p+\delta'
\end{equation}
the matrix
\begin{eqnarray}\label{Lambda}
\Lambda=\left(\begin{array}{cc}
\mu &\nu\\
\mu' &\nu'
\end{array}\right)
\end{eqnarray}
must satisfy the relation
\begin{eqnarray}
\Lambda\sigma\Lambda^T=\sigma;\quad 
\sigma=\left(\begin{array}{cc}
0&1\\
-1&0
\end{array}\right)\,.
\end{eqnarray}

Then, the marginal distribution is the diagonal matrix element
of the density operator in the transformed basis (\ref{Xeigen})
\begin{equation}\label{wrho}
w(x,\mu,\nu)=\langle x|\hat\rho|x\rangle={\rm Tr}
\{\hat\rho|x\rangle\langle x|\}
\end{equation}
or it is the diagonal matrix element in position representation of 
the transformed density
operator
\begin{equation}\label{wXS}
w(x,\mu,\nu)=\langle q|{\hat S}^{\dag}\hat\rho\hat S|q\rangle={\rm
Tr} \{{\hat S}^{\dag}\hat\rho\hat S|q\rangle\langle q|\}.
\end{equation}
The form of the shifted and squeezed operator $\hat S$ is well 
known \cite{Caves}. Choosing the
parameters $\mu=\cos\phi$ and 
$\nu=\sin\phi$, the operator $\hat S$ gives the marginal distribution for the
homodyne output of Ref. \cite{VogelRisken}.
In the case of $\mu=1$ and $\nu=0$ the marginal distribution is that for 
quadrature $\hat q$, i.e. $w(q,1,0)=\rho(q,q)=\langle q|\hat\rho|q\rangle$, 
while in the case of 
$\mu=0$ and $\nu=1$ the marginal distribution is that for the other
quadrature $\hat p$, i.e. $w(p,0,1)=\rho(p,p)=\langle p|\hat\rho|p\rangle$.

\section{Quantum evolution as a classical process}

We now derive the evolution equation for the marginal distribution function
$w$ using the invariant form of the connection between the marginal
distribution and the density operator given by the formula (\ref{rho}). 
Then, from the equation of
motion for the density operator which includes the interaction with 
environment $\chi(\rho)$ 
\begin{equation}\label{rhoH}
\partial_t\hat\rho=-i[\hat H,\hat\rho]+\chi(\rho)\,,
\end{equation}
we obtain the evolution equation for the marginal distribution in the form
\begin{equation}\label{wevo}
\int dxd\mu d\nu\; \left\{\dot w(x,\mu,\nu,t)\hat K_{\mu,\nu}+
w(x,\mu,\nu,t)\hat I_{\mu,\nu}\right\}=
\chi\left(\int dxd\mu d\nu\;w(x,\mu,\nu,t)\hat K_{\mu,\nu}\right)
\end{equation}
in which the known Hamiltonian determines the kernel $\hat I_{\mu,\nu}$ 
through the
commutator
\begin{equation}\label{I}
\hat I_{\mu,\nu}=i[\hat H,\hat K_{\mu,\nu}]\,,
\end{equation}
while the r.h.s. is functionally dependent on the marginal distribution.
The obtained integral-operator equation can be reduced to an 
integro-differential equation for
the function $w$ in some cases. Let us consider at first the 
situation in which $\chi(\rho)=0$,
the opposite situation will be discussed later.  Then, we 
represent the kernel operator
$\hat I_{\mu,\nu}$ in normal order form (i.e. all the momentum 
operators on the left side and the
position ones on the right side) containing the operator 
$\hat K_{\mu,\nu}$ as follow
\begin{equation}\label{Inormal}
:\hat I_{\mu,\nu}:={\cal R}(\hat p):\hat K_{\mu,\nu}:{\cal P}(\hat q)
\end{equation}
where ${\cal R}(\hat p)$ and ${\cal P}(\hat q)$ are, finite or 
infinite operator polynomials
(depending also on the parameters $\mu$ and $\nu$) determined 
by the Hamiltonian.
Then we calculate the matrix elements of the operator equation 
(\ref{wevo}) between the states 
$\langle p|$ and $|q\rangle$ obtaining
\begin{equation}\label{inteq}
\int dxd\mu d\nu\; \left\{\dot w(x,\mu,\nu,t)+
w(x,\mu,\nu,t){\cal R}(p){\cal P}(q)\right\}
\langle p|:\hat K_{\mu,\nu}:|q\rangle=0\,.
\end{equation}
If we suppose to write
\begin{equation}\label{RP}
{\cal R}(p){\cal P}(q)=\Pi(p,q)=\sum_n\sum_m c_{n,m}(z,\mu,\nu)p^nq^m\,,
\end{equation}
due to the particular form of the kernel in Eq. (\ref{K}), 
the Eq. (\ref{inteq}) can be rewritten
as
\begin{equation}\label{Piright}
\int dxd\mu d\nu\; \left\{\dot w(x,\mu,\nu,t)+
w(x,\mu,\nu,t){\overrightarrow\Pi}(\tilde p,\tilde q)\right\}
\langle p|:\hat K_{\mu,\nu}:|q\rangle=0\,,
\end{equation}
where $\tilde p$, $\tilde q$ are operators of the form
\begin{equation}\label{pqtilde}
\tilde p=\left(-\frac{i}{z}\frac{\partial}{\partial\nu}
+\frac{\mu}{2}z\right),\qquad
\tilde q=\left(-\frac{i}{z}\frac{\partial}{\partial\mu}
+\frac{\nu}{2}z\right);
\end{equation}
while $z$, in the space of variables $x,\mu,\nu$ should be 
intended as the derivative with
respect to $x$, i.e.
\begin{equation}\label{zx}
z\leftrightarrow i\frac{\partial}{\partial x}
\end{equation}
and when it appears in the denominator is understood as an 
integral operator.
Furthermore the right arrow over $\Pi$ means that, with respect 
to the order of Eq. (\ref{RP}),
the operators $\tilde p$ and $\tilde q$ act on the right, i.e. on 
$\langle p|:\hat K_{\mu,\nu}:|q\rangle$.
Under the hypothesis of regularity of $w$ on the boundaries, 
we can perform integrations by parts
in Eq. (\ref{Piright}) disregarding the surface terms, to get
\begin{equation}\label{Pileft}
\int dxd\mu d\nu\; \left\{\dot w(x,\mu,\nu,t)+
w(x,\mu,\nu,t){\overleftarrow\Pi}\left({\check p},{\check q}\right)\right\}
\langle p|:\hat K_{\mu,\nu}:|q\rangle=0\,,
\end{equation}
where now $\overleftarrow\Pi$ means that the operators $\check p$, $\check q$ 
\begin{equation}\label{tiltil}
{\check p}=\left(-\frac{i}{z}\frac{\partial}{\partial\nu}
-\frac{\mu}{2}z\right),\qquad
{\check q}=\left(-\frac{i}{z}\frac{\partial}{\partial\mu}
-\frac{\nu}{2}z\right);
\end{equation}
act on
the left, i.e. on the product of coefficients 
$c_{n,m}(-z,\mu,\nu)$ with the marginal distribution
$w$. Finally, using the completness property of the 
Fourier exponents given by
$\langle p|:\hat K_{\mu,\nu}:|q\rangle$
we arrive at the following equation of motion for the 
marginal distribution function
\begin{equation}\label{wmotion}
\partial_t w+w{\overleftarrow\Pi}\left({\check p},{\check q}\right)=0\,.
\end{equation}

Let us consider the  important
example of the motion of the particle in a potential with the Hamiltonian
\begin{equation}\label{HV}
\hat H=\frac{{\hat p}^2}{2}+V(\hat q);
\end{equation}
then the described procedure of calculating the normal 
order kernel (\ref{Inormal}) gives the
following form of the quantum dynamics in terms of a 
Fokker-Planck-like equation for the
marginal distribution
\begin{equation}\label{FPeq}
\dot w-\mu\frac{\partial}{\partial\nu}w
-i\left[V\left(\frac{-1}{\partial/\partial x}
\frac{\partial}{\partial\mu}-i\frac{\nu}{2}
\frac{\partial}{\partial x}\right)-
V\left(\frac{-1}{\partial/\partial x}
\frac{\partial}{\partial\mu}+i\frac{\nu}{2}
\frac{\partial}{\partial x}\right)\right]w=0
\end{equation}
which in the general case is an integro-differential equation. 
It is worth to remark that considering
the quadrature $X$ of Eq. (\ref{X}) to be dimensionless, the 
Planck constant $\hbar$, should
appears in the Eq. (\ref{FPeq}) to multiply the fisrt two terms. 
As a consequence it is clear
that the equation, even if classical-like, gives a quantum 
description of the system evolution
(as the Schr\"odinger equation).

Thus given a Hamiltonian  of the form (\ref{HV}) we can study 
the quantum evolution of
the system writing down a Fokker-Planck-like equation for 
the marginal distribution. Solving this
one for a given initial positive and normalized marginal 
distribution we can obtain the
quantum density operator $\hat\rho(t)$ according to Eq. (\ref{rho}). 
Conceptually it means that
we can discuss the system quantum evolution considering classical,
real positive and normalized
distributions for the measurable variable $X$ which is shifted, 
rotated and squeezed quadrature.
The distribution function which depends on extra parameters obeys 
a classical
equation which preserves the normalization condition of the
distribution. In this sense we always can reduce the quantum 
behaviour of the system to the
classical behaviour of the marginal distribution. Of
course, this statement respects the uncertainty relation because 
the measurable marginal
distribution is the distribution for one observable. That is the 
essential difference (despite of
some similarities) of the introduced marginal distribution from 
the discussed quasi-distributions,
including  the real positive Q-function, which depend on the two 
variables of the phase space and
are normalized with respect to these variables. We would point 
out that we do not derive
quantum mechanics from classical stochastic mechanics, i.e. we 
do not quantize any classical
stochastic process, our result is to present the quantum dynamics 
equations as classical
ones, and in doing this we need not only the classical Hamiltonian 
but also its quantum
counterpart.   

\section{Examples}

Let us choose as system to study a
driven harmonic oscillator of unit mass with an hamiltonian of the type
\begin{equation}\label{Hdriv}
H=\frac{p^2}{2}+\omega^2\frac{q^2}{2}-fq\,,
\end{equation}
then from Eq. (\ref{FPeq}) immediately follows
\begin{equation}\label{weqdriv}
\dot w-\mu\frac{\partial}{\partial\nu}w+\omega^2\nu
\frac{\partial}{\partial\mu}w
+f\nu\frac{\partial}{\partial x}w=0\,.
\end{equation}
Below we cosider solutions of some special cases of Eq. (\ref{weqdriv}), 
while the solution for the
complete equation will be given in a next section by using a propagator 
method.

\subsection{Free Motion}

For the free motion, $\omega=f=0$, the
evolution equation (\ref{weqdriv}) becomes the first order partial 
differential equation 
\begin{equation}\label{freeeq}
\dot w-\mu\frac{\partial}{\partial\nu}w=0\,,
\end{equation}
and it has a gaussian
solution of the form
\begin{equation}\label{freesolution}
w(x,\mu,\nu,t)=\frac{1}{\sqrt{2\pi\sigma_x(t)}}\exp\left\{-\frac{x^2}
{2\sigma_x(t)}\right\}
\end{equation}
where the dispersion of the observable $\hat x$ depends on time and 
parameters as follow
\begin{equation}\label{sigma}
\sigma_x(t)=\frac{1}{2}[\mu^2(1+t^2)+\nu^2+2\mu\nu t].
\end{equation}
The initial condition corresponds to the marginal distribution of 
the ground state of an
artificial harmonic oscillator calculated from the respective Wigner 
function \cite{Man1}.

\subsection{Harmonic Oscillator}

For the simple harmonic oscillator with frequency $\omega=1$, 
we have $f=0$ then Eq. (\ref{weqdriv})
becomes
\begin{equation}\label{hoeq}
\dot w-\mu\frac{\partial}{\partial\nu}w+\nu
\frac{\partial}{\partial\mu}w=0\,.
\end{equation}

If we consider the first excited state of the harmonic
oscillator, we know the Wigner function \cite{Gardiner}
\begin{equation}\label{W1}
W_1(q,p)=-2(1-2q^2-2p^2)\exp[-q^2-p^2].
\end{equation}
It results time independent due to the stationarity of the state, 
but for small $q$ and $p$ it
becomes negative while the solution of Eq. (\ref{hoeq})  
\begin{equation}\label{w1}
w_1(x,\mu,\nu,t)=\frac{2}{\sqrt{\pi}}[\mu^2+\nu^2]^{-\frac{3}{2}}x^2
\exp\left\{-\frac{x^2}{\mu^2+\nu^2}\right\}
\end{equation}
is itself time independent, but everywhere positive. 

Indeed, a time evolution is present explicitly in 
the coherent state, whose Wigner function is given by
\begin{equation}\label{Wc}
W_c(q,p)=2\exp\{-q^2-q_0^2-p^2-p_0^2+2(qq_0+pp_0)\cos t-(pq_0-qp_0)\sin t\}
\end{equation}
where $q_0$ and $p_0$ are the initial values of position and momentum.
For the same state the marginal distribution shows a more complicate 
evolution
\begin{eqnarray}\label{wc}
&w_c&(x,\mu,\nu,t)=\frac{1}{\sqrt{\pi}}[\mu^2+\nu^2]^{-\frac{1}{2}}\\
&\times&\exp\left\{-q_0^2-p_0^2-\frac{x^2}{\nu^2}
+2\frac{x}{\nu}(p_0\cos t-q_0\sin
t)\right\}\nonumber\\
&\times&\exp\left\{\frac{1}{\mu^2+\nu^2}\left[\frac{\mu}{\nu}x+
q_0(\mu\sin t+\nu\cos t)
+p_0(\nu\sin t-\mu\cos t)\right]^2\right\}.\nonumber
\end{eqnarray}

It is also interesting to consider the comparison between the Wigner 
function and the marginal
probability for non-classical states of the harmonic oscillator, 
such as female cat state
defined as \cite{Manko}
\begin{equation}\label{cat}
|\alpha_{-}\rangle=N_{-}(|\alpha\rangle-|-\alpha\rangle),
\quad\alpha=2^{-1/2}(q_0+ip_0)
\end{equation}
with 
\begin{equation}\label{N_}
N_{-}=\left\{\frac{\exp[(q_0^2+p_0^2)/2]}{4\sinh[(q_0^2+p_0^2)/2]}
\right\}^{\frac{1}{2}}
\end{equation}
and for which the Wigner function assumes the following form
\begin{eqnarray}\label{Wcat}
W_{-}(q,p)=2N_{-}^2e^{-q^2-p^2}&\{&e^{-q_0^2-p_0^2}\cosh[2(qq_0+pp_0)
\cos t+2(qp_0-pq_0)\sin t]
\nonumber\\
&-&\cos[2(qp_0-pq_0)\cos t-2(qq_0+pp_0)\sin t]\}. 
\end{eqnarray}
The corresponding marginal distribution is
\begin{eqnarray}\label{wcat}
w_{-}(x,\mu,\nu,t)&=&N_{-}^2[w_A(x,\mu,\nu,t)-w_B(x,\mu,\nu,t)
\nonumber\\
&-&w_B^*(x,\mu,\nu,t)+w_A(-x,\mu,\nu,t)]
\end{eqnarray}
with
\begin{eqnarray}
&w_A&(x,\mu,\nu,t)=\frac{1}{\sqrt{\pi}}[\mu^2+\nu^2]^{-\frac{1}{2}}\\
&\times&\exp\left\{-q_0^2-p_0^2-\frac{x^2}{\nu^2}+2\frac{x}{\nu}
(p_0\cos t-q_0\sin
t)\right\}\nonumber\\
&\times&\exp\left\{\frac{1}{\mu^2+\nu^2}\left[\frac{\mu}{\nu}x+
q_0(\mu\sin t+\nu\cos t)
+p_0(\nu\sin t-\mu\cos t)\right]^2\right\}\nonumber
\end{eqnarray}
and
\begin{eqnarray}
&w_B&(x,\mu,\nu,t)=\frac{1}{\sqrt{\pi}}[\mu^2+\nu^2]^{-\frac{1}{2}}\\
&\times&\exp\left\{-\frac{x^2}{\nu^2}
-2i\frac{x}{\nu}(q_0\cos t+p_0\sin
t)\right\}\nonumber\\
&\times&\exp\left\{\frac{-1}{\mu^2+\nu^2}\left[-i\frac{\mu}{\nu}x+
q_0(\mu\cos t-\nu\sin t)
+p_0(\mu\sin t+\nu\cos t)\right]^2\right\}.\nonumber
\end{eqnarray}

The presented examples show that, for the evolution of the state of 
a quantum system, one
could always associate the evolution of the probability density 
for the random classical variable
$X$ which obeys "classical" Fokker-Planck-like equation, and this 
probability density contains the
same information (about a quantum system) which is contained in 
any quasi-distribution function. But
the probability density has the advantage to behave completly as 
the usual classical one. The
physical meaning of the "classical" random variable $X$ is 
transparent, it is considered as the
position in an ensemble of shifted, rotated and scaled rest 
frames in the classical phase
space of the system under study. We could remark that for 
non normalized quantum states, like the
states with fixed momentum (De Broglie wave) or with fixed position,
the introduced map in Eq. (\ref{rho}) may be  preserved. 
In this context the plane
wave states of free motion have the marginal distribution 
corresponding to the classical white
noise as we shall see below.

\subsection{Squeezed Coherent States}

Here we will consider the marginal distribution $w(x,\mu,\nu)$ 
for the squeezed
coherent states of the harmonic oscillator. Since the Wigner 
function of these pure gaussian
states may be represented in the form \cite{Dodonov183}
\begin{equation}\label{Wsq}
W_{\alpha}(q,p)=2\exp\left[-\frac{1}{2}\left(p-{\overline p}(t),
q-{\overline q}(t)\right)\,
{\bf m}^{-1}
\left(\begin{array}{c}p-{\overline p}(t)\\
q-{\overline q}(t)\end{array}\right)
\right]\,,
\end{equation}
where $\bf m$ is the dispersion matrix
\begin{equation}\label{m}
{\bf m}=
\left(\begin{array}{cc}
 \sigma_p(t)&\sigma_{pq}(t)\\
 \sigma_{pq}(t)&\sigma_q(t)
\end{array}\right)
\end{equation}
and ${\overline p}(t)$ and ${\overline q}(t)$ are 
the mean values of the quadratures 
\begin{equation}\label{qpave}
{\overline q}(t)=\sqrt{2}\Re(\alpha e^{-it}),\quad
{\overline p}(t)=\sqrt{2}\Im(\alpha e^{-it}).
\end{equation}
The variances in eq. (\ref{m}) are given by
\begin{eqnarray}\label{var}
\sigma_p(t)&=&\frac{1}{2}\left(s\cos^2t+\frac{1}{s}\sin^2t\right)\,,\\
\sigma_q(t)&=&\frac{1}{2}\left(\frac{1}{s}\cos^2t+s\sin^2t\right)\,,\\
\sigma_p(t)&=&\frac{1}{2}\left(s-\frac{1}{s}\right)\sin t\cos t\,,
\end{eqnarray}
with $s$ the squeezing parameter. Using Eq. (\ref{Wsq}) 
in the formulae (\ref{W}) and
(\ref{wF}), we obtain for the marginal distribution the expression 
\begin{equation}\label{wsq}
w(x,\mu,\nu,t)=\frac{1}{\sqrt{2\pi\sigma_x(t)}}
\exp\left\{-\frac{[x-\mu{\overline q}(t)
-\nu{\overline p}(t)]^2}{2\sigma_x(t)}\right\}\,,
\end{equation}
where 
\begin{equation}\label{sigmasq}
\sigma_x(t)=\frac{1}{2}[\mu^2\sigma_q(t)+\nu^2\sigma_p(t)
+2\mu\nu\sigma_{pq}(t)]\,.
\end{equation}
Let us now take the limit $s\to 0$, this means that our 
marginal distribution becomes a delta
function 
\begin{equation}\label{wdelta}
\lim_{s\to 0}w=\delta\left(x-(\mu q_0+\nu p_0)\cos t
-(\mu p_0-\nu q_0)\sin t\right)\,,
\end{equation}
and as a consequence its spectrum will be constant and equal 
to unity for each values of the
variable conjugate to $X$, thus it will correspond to the 
white noise spectrum. On the other
hand, the nonnormalized quantum states, like the states with 
fixed momentum (De Broglie wave) or
with fixed position, have a marginal distribution normalized 
and everywhere equal to one. Thus
plane wave states of free motion correspond to the classical 
white noise distribution. 

\section{Evolution in the Presence of Environmental Interaction}

When a system is coupled with the "rest of Universe" the time 
evolution of the density 
operator is no longer unitary, and to treat the problem at quantum 
level, one needs of 
some approximations; usually the starting point is a simple system 
as an harmonic 
oscillator which linearly interacts with a bath idealized as an 
infinity of other harmonic 
oscillators, then the (master) equation for the density operator 
becomes \cite{Lou}
\begin{eqnarray}\label{master eq}
\dot\rho&=&-i[a^{\dag}a,\rho]+\chi(\rho)\nonumber\\
\chi(\rho)&=&\frac{\gamma}{2}({\overline n}+1)(2a\rho a^{\dag}
-a^{\dag}a\rho- 
\rho a^{\dag}a)\nonumber\\
&+&\frac{\gamma}{2}{\overline n}(2a^{\dag}\rho a- aa^{\dag}\rho
-\rho aa^{\dag})
\end{eqnarray}
where $\gamma$ is the damping constant characterizing the relaxation 
time of the system, 
$a,\,a^{\dag}$ are the boson operators of the system and $\overline n$ 
is the number of the 
thermal excitation of the bath.
Using Eq. (\ref{rhoH}) in the interaction picture and performing step 
by step the same 
procedure that leads to Eq. (\ref{wmotion}), one may describes the 
damped evolution by
 means of
\begin{equation}\label{weqdamp}
\dot w=\frac{\gamma}{2}\left[2-\frac{\partial}{\partial\nu}\nu
-\frac{\partial}{\partial\mu}\mu
+\frac{1}{2}(\mu^2+\nu^2)\frac{\partial^2}{\partial x^2}\right]w\,,
\end{equation}
where we have assumed for simplicity ${\overline n} =0$, a situation 
common in quantum 
optical systems.
In Eq. (\ref{weqdamp}) we recognize the Fokker-Planck equation where 
the diffusion 
term is given by the proper stochastic term  while the drift by the 
parameters (the factor 2 
can be eliminated by a simple transformation $w=\tilde w e^{\gamma t}$).
The solution of Eq. (\ref{weqdamp}), with coherent initial 
excitation $q_0,\,p_0$, is
\begin{equation}\label{wdampsol}
w(x,\mu,\nu,t)=\frac{1}{\pi}\frac{1}{\sqrt{\pi(\mu^2+\nu^2)}}\exp\left\{-
\frac{\left[x+(\mu q_0-\nu p_0)
e^{-\gamma t/2}\right]^2}{\mu^2+\nu^2}\right\}
\end{equation}
which is exactely the Fourier transform of the Wigner function 
for the damped harmonic 
oscillator given by \cite{Gardiner}
\begin{equation}\label{Wdamp}
W(q,p)=2\exp\left[-(q-q_0e^{-\gamma t/2})^2
-(p-p_0e^{-\gamma t/2})^2\right]\,.
\end{equation}
This is a proof that the developed formalism is consistent also 
in the case of open quantum
systems.

\section{Quantum measurements and classical measurements}
 
In this section we will discuss the concept of quantum measurements 
in the frame of the developed
approach. It is a well known steatment \cite{WZ}, \cite{Bell} that 
quantum mechanics suffers
from an inconsistence in the sense that it needs, for its understanding, 
of a classical device
measuring quantum observables. Due to this the theory of measurements 
suppose that there exist two
worlds: the classical one and the quantum one. Of course in the 
classical world the measurements of
classical observables are produced by classical devices. 
In quantum world the measurements of
quantum observables are produced by classical devices too. 
Due to this the theory of quantum measurements is considered as
something very specifically different from the classical measurements.

Recentely it has been proposed some schemes \cite{Per}, \cite{Home} 
to resolve the dichotomy
between the measured microsystem and the measuring macroapparatus, however
it is phsycologically accepted that to understand the physical 
meaning of a measurement in classical
world is much easier than to understand the physical meaning of an analogous
measurement in quantum world. 

Our aim is to show that in fact all the roots of difficulties of 
quantum measurements are present
in the classical measurements as well. Using the invertible map, 
of the quantum states (both
normalized and nonnormalized) and classical states (described by 
classical distributions-generalized
functions), given by Eq. (\ref{rho}) we could conclude that the 
complete information about
a quantum state is obtained from purely classical measurements 
of the position of a particle, made 
by classical devices in each reference frame of the ensamble of 
the classical reference frames, 
which are shifted, scaled and rotated in the classical phase space.

These measurements do not need of any quantum language, if we know 
how to produce, in the
classical world (using the notion of classical position and momentum), 
reference frames in
the classical phase space differing from each other by rotation, 
scaling and shifting of the
axis of the reference frame and how to measure only the position 
of the particle from the point of
view of these different reference frames. Thus, knowing how to 
obtain the classical marginal
distribution function
$w(x,\mu,\nu)$ which depends on the parameters $\mu,\nu,\delta$, 
labeling each
reference frame in the classical phase space, we reconstruct 
through the map (\ref{rho}) the quantum
density operator. 

By this approach, we avoid the unpleasant paradox of quantum 
world which needs for its explanation
measurements by a classical apparatus. Nevertheless all the 
difficulties of the quantum approach
continue to be present, but in a different classical form. 
In fact, if we consider for example the notion of wave function 
collapse \cite{Von}, it is
displaced in the classical framework, since if we idealize the 
measuring apparatus as a bath with
which the system interacts \cite{CL}, then a reduction of the 
probability distribution (as our
marginal distribution) occours as soon as we "pick" a value 
(hence a trajectory) of the classical
stochastic process associated to the observable 
(as that of Eq. (\ref{X})).

About the developped formalism, we are aware that
the crucial point might be the practical realization of the 
generic linear quadratures such as in
Eq. (\ref{X}). Then, let us consider a practical implementation, 
in the optical domain.
The quadrature of Eq.
(\ref{X}) could be experimentally accessible by using for example 
the squeezing
pre-amplification (pre-attenuation) of a field mode which is going 
to be measured (a similar method
in different context was discussed in Ref. \cite{leopauPRL}). 
In fact, let
$\hat a$ be the signal field mode to be detected, when it passes 
through a squeezer it becomes
${\hat a}_s={\hat a}\cosh s-{\hat a}^{\dag}e^{i\theta}\sinh s$, 
where $s$ and $\theta$ characterize
the complex squeezing parameter
$\zeta=se^{i\theta}$
\cite{LK}. Then, if we subsequently detect the field by 
using the balanced homodyne scheme, we get
an output signal proportiopnal to the average of the following quadrature
\begin{equation}\label{N1-N2}
{\hat E}(\phi)=\frac{1}{\sqrt{2}}({\hat a}_se^{-i\phi}+{\hat
a}^{\dag}_se^{i\phi})\,,
\end{equation}
where $\phi$ is the local oscillator phase. When this phase is locked to
that of the squeezer, such that $\phi=\theta/2$, Eq. (\ref{N1-N2}) becomes
\begin{equation}\label{quamea}
{\hat E}(\phi)=\frac{1}{\sqrt{2}}\left({\hat a}
e^{-i\theta/2}[\cosh s-\sinh s]
+{\hat a}^{\dag}e^{i\theta/2}[\cosh s-\sinh s]\right)\,,
\end{equation}
which, essentially, coincides with Eq. (\ref{X}), if one recognizes 
the independent parameters
\begin{equation}\label{munumeas}
\mu=[\cosh s-\sinh s]\cos(\theta/2);\quad\nu=[\cosh s-\sinh s]
\sin(\theta/2)\,.
\end{equation}
The shift parameter $\delta$ has not a real physical meaning, 
since it causes only a displacement of
the distribution along the $X$ line without changing its shape, 
as can be evicted from Eqs.
(\ref{wX}) and (\ref{w}). So, in a practical situation it can be omitted.
To be more precise, the shift parameter does not play a real 
physical role in the measurement 
process,
it has been introduced for formal completeness and it expresses 
the possibility to achieve the 
desired
marginal distribution by performing the measurements in an 
ensemble of frames which are  each
other shifted; (related method was early discussed in Ref.
\cite{Royer}). In an electro-optical system this only means 
to have the freedom of using different
photocurrent scales in which the zero is shifted by a known amount.

\section{Connection with measurements in homodyne tomography}

At this point,
a comparision with the usual tomographic technique, used in 
the experiments of the
type of Ref. \cite{Raymer}, is useful. To this end we recall 
that in this case the timelike
evolution of the system is brought about by the parameters changing, 
thus no explicit time
dependence of
$w$ is needed. Furthermore, we note that a relation betweeen the 
density operator and the marginal
distribution analogous to Eq. (\ref{rho}) can be derived starting 
from another operator identity
such as
\cite{CahillGlauber}
\begin{equation}\label{rho2}
\hat\rho=\int \frac{d^2\alpha}{\pi}\hbox{Tr}\{\hat\rho\hat D 
(\alpha)\}D^{-1}(\alpha)
\end{equation}
which, by the change of variables $\mu=-\sqrt{2}\Im \,\alpha,
\quad\nu=\sqrt{2}\Re \,\alpha$,
becomes
\begin{equation}\label{rho3}
\hat\rho=\frac{1}{2\pi}\int d\mu d\nu\quad \hbox{Tr}\{\hat\rho 
e^{-i\hat X}\}e^{i\hat X}
=\frac{1}{2\pi}\int d\mu d\nu\quad \hbox{Tr}\{\hat\rho 
e^{-i\hat x}\}e^{i\hat x}\,.
\end{equation}
The trace can be now evaluated
using the complete set of eigenvectors $\{|x\rangle\}$ for the operator
$\hat x$, obtaining
\begin{equation}\label{trace}
\hbox{Tr}\{\hat\rho e^{-i\hat x}\}=\int dx\quad w(x,\mu,\nu)e^{-ix}
\end{equation}
then, putting this one into Eq. (\ref{rho3}), we have a relation 
of the same form of Eq.
(\ref{rho}) with the kernel given by
\begin{equation}\label{kerbis}
\hat K_{\mu,\nu}=\frac{1}{2\pi}e^{-ix}e^{i\hat x}=
\frac{1}{2\pi}e^{-ix}e^{i\mu{\hat q}+i\nu{\hat p}},
\end{equation}
which is the same of Eq. (\ref{K}) setting $z=1$. It means 
that we now have 
only one particular Fourier component due to the particular 
change of variables (the most general
should be $z\mu=-\sqrt{2}\Im \,\alpha$ and $z\nu=\sqrt{2}\Re \,\alpha$).

In order to reconstruct the usual tomographic formula for the 
homodyne detection \cite{DarPaul} we
need to pass in polar variables, i.e. $\mu=-r\cos\phi,\quad\nu=-r\sin\phi$,
then 
\begin{equation}
\hat x\to -r\hat x_{\phi}=-r[\hat q\cos\phi+\hat p\sin\phi].
\end{equation}
Furthermore, indicating with $x_{\phi}$ the eigenvalues of the 
quadrature $\hat x_{\phi}$, we have
\begin{equation}
\hbox{Tr}\{\hat\rho e^{-i\hat x}\}=\hbox{Tr}\{\hat\rho 
e^{ir\hat x_{\phi}}\}=
\int dx_{\phi}\quad w(x_{\phi},\phi)e^{irx_{\phi}}
\end{equation}
and thus, from Eq. (\ref{rho3})
\begin{equation}
\hat\rho=\int d\phi dx_{\phi}\quad w(x_{\phi},\phi)\hat K_{\phi}
\end{equation}
with
\begin{equation}\label{kerAPL}
\hat K_{\phi}=\frac{1}{2\pi}\int dr\quad re^{ir(x_{\phi}-\hat x_{\phi})}
\end{equation}
which is the same of Ref. \cite{DarPaul}.
Substantially, the kernel of Eq. (\ref{kerAPL}) is given by 
the radial integral of the kernel of 
Eq.
(\ref{kerbis}), and this is due to the fact that we go from 
a general transformation, with two 
free
parameters, to a particular transformation (homodyne rotation) 
with only one free parameter, and 
then
we need to integrate over the other one. 
This derivation follows Ref. \cite{Man2}.

\section{Generating function for momenta}

Since the marginal distribution $w(x,\mu,\nu)$ has all the 
properties of the classical
probability density, one could calculate highest momenta 
for the shifted and sqeezed quadrature
$\hat x$. We have by definition
\begin{equation}\label{mom}
\langle{\hat x}^n\rangle=\int x^n\,w(x,\mu,\nu)dx,
\end{equation}
thus for the mean value $(n=1)$
\begin{equation}\label{ave}
\langle\hat x\rangle=\int x\,w(x,\mu,\nu)dx,
\end{equation}
and for the quadrature dispersion one has
\begin{equation}\label{sigma}
\sigma_x=\langle {\hat x}^2\rangle-\langle \hat x\rangle^2=
\int x^2\,w(x,\mu,\nu)dx-\left[\int x\,w(x,\mu,\nu)dx\right]^2.
\end{equation}
As in the standard probability theory \cite{Gne}, to calculate 
highest momenta for the shifted and
squeezed quadrature one could introduces the generating function
\begin{equation}\label{G}
G(i\lambda)=\sum_{n=0}^{\infty} \frac{(i\lambda)^n}{n!}
\langle{\hat x}^n\rangle.
\end{equation}
Then the highest momenta are the coefficients of the Taylor 
series for the decomposition of the
generating function with respect to the parameter $(i\lambda)$. 
We will express this generating
function in terms of the Wigner function for the quantum system. 
Inserting Eq. (\ref{w})
into Eq. (\ref{mom}) we have
\begin{equation}\label{momwig}
\langle{\hat x}^n\rangle=
\int x^n e^{-ik(x-\mu q-\nu p)}W(q,p)\frac{dqdpdkdx}{(2\pi)^2}
\end{equation}
and inserting this one into Eq. (\ref{G}) we arrive at
\begin{equation}\label{Gwig}
G(i\lambda)=
\int e^{-ik(x-\mu q-\nu p)+i\lambda x}W(q,p)\frac{dqdpdkdx}{(2\pi)^2}.
\end{equation}
Now integrating, first over the quadrature variable $x$ and 
then over the variable $k$, we get
\begin{equation}\label{gen}
G(i\lambda)=
\int \frac{dqdp}{2\pi}W(q,p)e^{i\lambda (\mu q+\nu p)}.
\end{equation}
This expression shows that the generating function for the 
quadrature highest momenta is
determined by the Fourier components of the system Wigner function
\begin{equation}\label{WF}
W_F(a,b)=\frac{1}{(2\pi)^2}\int W(q,p)e^{iqa+ipb}dqdp
\end{equation}
i. e.
\begin{equation}\label{GWF}
G(i\lambda)=\frac{1}{2\pi}W_F(\lambda\mu,\lambda\nu).
\end{equation}
Thus, having the Wigner function of the quantum state and 
calculating its Fourier component, we
may determine the generating function which depends on the 
extra parameters $\mu,\,\nu$.
On the other hand, since from Eq. (\ref{WF}) we have the 
inverse Fourier transform
\begin{equation}\label{WWF}
W(q,p)=\frac{1}{(2\pi)^2}\int W_F(a,b)e^{-iqa-ipb}dadb
\end{equation}
we can express the Wigner function trough the generating function as
\begin{equation}\label{WG}
W(q,p)=\frac{1}{(2\pi)^3}\int e^{-i\mu q-i\nu p}G(i)d\mu d\nu,
\end{equation}
where we have taken $\lambda=1$ and integrated over the parameters 
$\mu$ and $\nu$ on which the
generating function depends.

Hence we conclude that the quantum information about the state 
is completely contained in the
expression for the generating function. It reflects the fact 
that measuring the shifted, rotated
and squeezed quadrature we measure the momenta of the marginal 
distribution
$w(x,\mu,\nu)$, and in fact we could reconstruct the generating 
function as a
function of the extra parameters $\mu,\nu$. Thus, the Wigner 
function of the
system is obtained from Eq. (\ref{WG}).

\section{Conditional Probability}

The direct extension of classical probability concepts leads also 
to the conditional probability
notion.  Using the convention that $\bf x$ means the vector given 
by the quadrature variable $x$ and
the  parameters $\mu$ and $\nu$, the joint probability 
$w({\bf x}_1,t_1;{\bf x}_2,t_2)$
is defined as the probability to have $x_1$ as result of the 
quadrature measurement at time $t_1$ 
in the frame $\{\mu_1,\nu_1\}$ {\it and}  $x_2$ as result of 
the quadrature measurement at time 
$t_2$ in the frame $\{\mu_2,\nu_2\}$. Then the conditional 
probability follows as
\begin{equation}\label{wcond}
w({\bf x}_1,t_1|{\bf x}_2,t_2)=
\frac{ w({\bf x}_1,t_1;{\bf x}_2,t_2)}{ w({\bf x}_2,t_2)}\,.
\end{equation}
As a consequence the Chapman-Kolmogorov equation \cite{HSM} 
will be satisfied, i.e.
\begin{equation}\label{smol}
w({\bf x}_1,t_1)=\int d^3{\bf x}_2 w({\bf x}_1,t_1|{\bf x}_2,t_2) 
w({\bf x}_2,t_2)\,,
\end{equation}
so that the defined conditional probability can be interpreted 
as the propagator for the marginal 
distribution. 
The physical meaning of the real positive propagator (\ref{wcond}) 
is the following: it is
the transition probability to go from the position $x_2$ in which 
the particle is situated at initial
time $t_2$ in the reference frame labeled by scaling and rotation 
parameters $\{\mu_2,\nu_2\}$,
into the position $x_1$ at the moment
$t_1$ in the reference frame labeled by the parameters $\{\mu_1,\nu_1\}$.

We would remark that, even thought the stochastic process on which the 
marginal distribution depends is only one, we need to integrate also on 
the variables representing
the  parameters since the same process may "come" from different frames.
In fact really the normalization condition, as consequence of 
Eqs. (\ref{norm}) and (\ref{smol}),
can be read as 
\begin{equation}\label{doublenorm}
\int dx_1dx_2\; w({\bf x}_1,t_1|{\bf x}_2,t_2) w({\bf x}_2,t_2)=1\,.
\end{equation}

In order to see the equation at which the conditional probability 
(\ref{wcond}) obeys, we insert  
Eq. (\ref{smol}) into Eq. (\ref{wmotion}), obtaining
\begin{equation}\label{CKeq}
w({\bf x}_1,t_1|{\bf x}_2,t_2)\overleftarrow{\left(\partial_{t_1}
+\Pi(\check p,\check q)\right)}
=\delta^3({\bf x}_1-{\bf x}_2)\delta(t_1-t_2)\,,
\end{equation}
that is the analogous of the differential Chapman-Kolmogorov 
equation \cite{HSM}.

As an application, let us consider the case of the driven 
harmonic oscillator for which, from Eq.
(\ref{weqdriv}), we have
\begin{equation}\label{dhocondeq}
\left(\frac{\partial}{\partial t_1}-\mu_1\frac{\partial}{\partial\nu_1}
+\omega^2\nu_1\frac{\partial}{\partial\mu_1}+f\nu_1
\frac{\partial}{\partial x_1}
\right)
w({\bf x}_1,t_1|{\bf x}_2,t_2) =\delta^3({\bf x}_1-{\bf x}_2)
\delta(t_1-t_2)\,,
\end{equation}
whose solution, for $t_1>t_2$, will be
\begin{eqnarray}\label{dhocondsol}
w({\bf x}_1,t_1|{\bf x}_2,t_2) &=&
\delta\left(\nu_2-\mu_1\sin[\omega(t_1-t_2)]-\omega\nu_1
\cos[\omega(t_1-t_2)]\right)\nonumber\\
&\times&\delta\left(\mu_1\cos[\omega(t_1-t_2)]-\omega\nu_1
\cos[\omega(t_1-t_2)]-\mu_2\right)
\nonumber\\ 
&\times&\delta\left(x_1-x_2-\mu_1\frac{f}{\omega^2}
\{1-\cos[\omega(t_1-t_2)]\}
-\nu_1\frac{f}{\omega}\sin[\omega(t_1-t_2)]\right) \,.\nonumber\\
\end{eqnarray}

Now, by means of Eqs. (\ref{smol}) and (\ref{dhocondsol}) we may 
derive the solution of Eq.
(\ref{weqdriv}) starting for example from an initial coherent 
condition characterized by $q_0$ and
$p_0$, i.e. Eq. (\ref{wc}) at $t=0$, obtaining 
\begin{eqnarray}\label{wcdriv}
&&w(x,\mu,\nu)=\frac{1}{\sqrt{\pi(\mu^2+\omega^2\nu^2)}}\nonumber\\
&&\times\exp\Bigg\{-
\Bigg[x-\mu\frac{f}{\omega^2}(1-\cos(\omega t))-\nu
\frac{f}{\omega}\sin(\omega t)
-q_0(\mu\cos(\omega t)-\omega\nu\sin(\omega t))\nonumber\\
&&+p_0(\mu\sin(\omega t)+\omega\nu\cos(\omega t))\Bigg]^2\Bigg/
(\mu^2+\omega^2\nu^2)\Bigg\}\,,
\end{eqnarray}
where we have taken ${\bf x}_1={\bf x}$, $t_1=t$ and $t_2=0$.
Of course, if we set $f=0$ and $\omega=1$ in Eq. (\ref{wcdriv}) 
we have again the solution
(\ref{wc}).  

Finally, as special case of the propagator formula (\ref{smol}), 
we can consider the time 
evolution of the marginal distribution of Ref. \cite{VogelRisken}
\begin{eqnarray}\label{VRevol}
w(x_1,\mu_1=\cos\phi,\nu_1=\sin\phi,t_1)&=&\nonumber\\
\int &d^3{\bf x}_2& 
w(x_1,\mu_1=\cos\phi,\nu_1=\sin\phi,t_1|{\bf x}_2,t_2) 
w({\bf x}_2,t_2)\,.\nonumber\\
\end{eqnarray}
This could be useful as a connection between our formalism 
and the homodyne tomography at different
times.

\section{Conclusions}

We have shown that it is possible to bring the quantum dynamics back 
to the classical description in
terms of a probability distribution containing (over)complete information. 
The time evolution of
a measurable probability for the discussed observables could be useful 
both for the prediction of
the experimental outcomes  at a given time and, as mentioned above, 
to achieve the quantum state
of the system at any time.
Furthermore, the symplectic transformation of Eq. (\ref{X}) could 
be represented as a composition
of shift, rotation and squeezing.
So, we would emphasize that our procedure allows to transform the 
problem of quantum measurements
(at least for some observables) into a problem of classical  
measurements with an ensemble of
shifted, rotated and scaled reference frames in the (classical) phase space.

Quite generally physics distinguishes between the dynamical 
law and the state of a system. The
state contains the complete statistical information about an 
ensemble of physical objects at a
particular moment, while the dynamical law determines the change 
of the status quo at the next
istant of time.  But can we use the dynamical law to infer the 
state (for example) of a moving
particle after position measurements have been performed?
For istance, in molecular emission tomography \cite{Walmsley} 
the quantum state of a mulecular
vibration has been determined from its elongation encoded 
in the time-evolved fluorescence
spectrum, while the usual standard tomography schemes \cite{Raymer} 
have been restricted to harmonic
oscillators or free particles for which one has a simple shearing 
or rotation in the phase space;
however the developed formalism is able to
infer the state of a particle moving in an arbitrary potential 
\cite{Olga} provided to have
positionlike measurements in different frames (an analogous 
problem using nontomographic approach
has been studied in Ref. \cite{LR}).
Of course, in some situations the measurements of instantaneous 
values of the
marginal distribution for different values of the parameters 
could be replaced by measuring the
distribution for these parameters which change in time. Such 
measurements may be
consistent with the system evolution if the time variation 
of parameters is much faster than the
natural evolution of the system itself. In this case the 
state of the system does not change during
the measurement process and one obtains the instantaneous 
value of the marginal distribution and
that of the Wigner function. 

Finally, we believe that our "classical" approach could be 
a powerful tool to
investigate complex quantum systems as for example chaotic 
systems in which the quantum chaos could
be considered in a frame of equations for a real and positive 
distribution function.
On the other hand, since the symplectic transformations are 
usually involved in the theory of
special relativity, we could think to apply the developped 
formalism for a relativistic
formulation of the quantum measurement theory. These will 
be the subjects of future papers.

\section*{\bf Acknowledgments}

This work has been partially supported by European Community under the 
Human Capital and Mobility (HCM) programme and by 
Istituto Nazionale di Fisica Nucleare. V. I. M.
would also like to acknowledge the University of 
Camerino for kind hospitality.

\end{document}